\documentclass[12pt,a4paper]{article}

\usepackage{amsbsy,latexsym,amssymb}
\usepackage{graphicx}
\usepackage{color}

\newcommand{\R}{\mathfrak{R}}

\newcommand{\kb}{{\boldsymbol k}}

\begin{document}

\begin{titlepage}

\vskip19truemm
\begin{center}{\Large{\textbf{Soft Collinear Degeneracies in an
Asymptotically Free Theory}}}\\ [12truemm] \textsc{Martin
Lavelle$^a$}\footnote{email: mlavelle@plymouth.ac.uk}, \textsc{David McMullan$^a$}\footnote{email: dmcmullan@plymouth.ac.uk} and \textsc{Tom Steele$^b$}\footnote{email: Tom.Steele@usask.ca} \\[5truemm]
\textit{$^a$School of Computing and Mathematics\\
The University of Plymouth\\
Plymouth, PL4 8AA\\ UK} \\[3truemm]
\textit{$^b$Department of Physics and Engineering Physics,\\ University of Saskatchewan,\\ Saskatoon,\\ Saskatchewan, S7N 5E2\\ Canada }
\end{center}

\bigskip\bigskip\bigskip
\begin{quote}
\textbf{Abstract} In asymptotically free theories with collinear divergences it is sometimes claimed that these divergences cancel if  one  sums over initial and final state degenerate cross-sections and uses an off-shell renormalisation scheme. We show for scalar $\phi^3$ theory in six dimensions that there are further classes of soft collinear divergences and that they do not cancel. Furthermore, they yield a non-convergent series of terms at a fixed order of perturbation theory. Similar effects in gauge theories are also summarised.
\end{quote}

\end{titlepage}

\section*{Introduction}

\noindent Infra-red (IR) divergences plague the extraction of physical cross-sections in gauge theories. In practice most calculations are restricted to infra-red safe quantities. However, according to the Lee-Nauenberg (LN) theorem~\cite{Lee:1964is}, by summing over all degenerate states (including \emph{both} final and initial state radiation) one should obtain an IR finite (sufficiently inclusive) cross-section. This should be contrasted with the Bloch-Nordsieck (BN) trick~\cite{Bloch:1937pw} where one just sums over final state radiation. It should be noted that the BN trick breaks down in theories with collinear divergences, e.g., QCD or QED with massless fermions.\footnote{It is worth noting that one can have large collinear-type logarithms even for massive particles in some kinematic regions. These should also be removed by a proper treatment of the infra-red.}

The indistinguishable processes which are summed over depend upon experimental resolutions. In this paper we will consider such two thresholds. There is an energy resolution, $\Delta$, such that particles with an energy less than this cannot be detected. Such undetected particles are called soft. There is also an angular resolution, $\delta$, such that \emph{massless} particles within a cone cannot be  distinguished from each other. Note that such collinear particles are not necessarily soft, they can carry a significant fraction of the energy in the jet, i.e.,  have energy greater than $\Delta$, in which case they are called semi-hard. We will regulate collinear divergences via a small mass, $m$.

We should also emphasise that, following on the work of Yennie and collaborators~\cite{Yennie:1961ad} it is not allowed to set the energy resolution to zero. They argued that the inclusive cross-section vanishes as the resolution is taken to zero. It is also, of course, the case that any real experiment has a finite resolution inherent to its detectors, cuts etc.

It is sometimes argued that soft divergences may be removed solely by summing over the soft emissions (BN) while the collinear divergences are separately cancelled by summing over emission and absorbtion processes. If one includes semi-hard absorbtion, but not soft absorbtion, one is left with $\ln(m)$-singularities multiplied by powers or logarithms of $\Delta$. These singularities are cancelled by including (non-eikonal) terms in soft absorbtion processes. These terms are soft-finite by power counting. However, the eikonal terms in these same soft absorbtion diagrams introduce fresh soft divergences. This indicates that, in the spirit of the LN~theorem, one needs to sum over all degenerate states for soft divergences~\cite{Lavelle:2005bt,Lavelle:2006vv}.

Below we will investigate massless scalar $\phi^3$ theory in $D=6$ which has been often used as a toy model of QCD.  This is because the theory is asymptotically free. It also suffers from collinear divergences although, as a consequence of the higher dimensionality, it does not have soft divergences. It thus offers a cleaner testing ground for studies of collinear divergences than, say, massless QED where there are overlapping soft and collinear singularities. We will first recap how the collinear singularities manifest themselves and how the LN theorem breaks down in the on-shell renormalisation scheme. In this asymptotically free theory it has been argued, see, e.g.,~\cite{sred:2007}, that a different renormalisation scheme should be used and that this leads to collinear finite inclusive cross-sections. We will show, however, that previous arguments neglect a new source of degeneracies which yield additional collinear divergences. These divergences, as we will show, are also present in gauge theories. We finally discuss some alternative approaches.

\section*{Scattering at One Loop}

We will study the process of two particle elastic scattering in
$D=6$ scalar field theory with an  intereaction term of $\lambda \phi^3/6$.   This theory offers a simple model for some of the phenomena of QCD: in particular it is asymptotically free and exhibits the collinear divergences which are central to  QCD.

At one loop diagrams such as
\begin{center}
\includegraphics[width=2.5cm,angle=90]{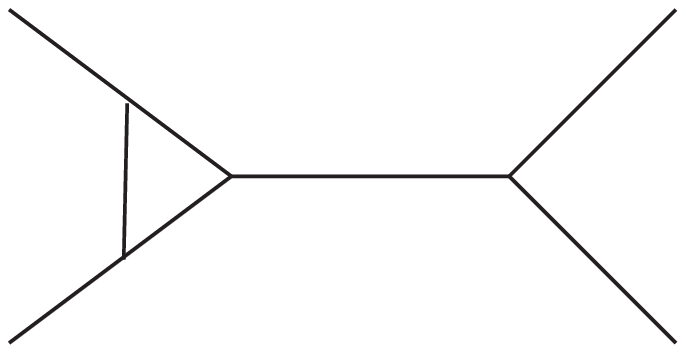}
\end{center}
generate collinear divergences in different channels.  There are also UV divergences, which we will ignore. It is important to note that, due to the large number of dimensions, there are no soft divergences in this theory. We will regularise the collinear divergences by giving the fields a small mass $m$, the collinear divergences then appearing as $\ln(m)$ factors.

In the context of this model, the  Bloch-Nordsieck trick,  summing over final states, does not lead to a collinear finite cross-section and  thus one is naturally led to  follow Lee and Nauenberg and sum over initial and final states. This introduces diagrams like
\[
\begin{array}{cc}
  \begin{minipage}{.5\textwidth}
 \begin{center}
 \includegraphics[width=2.5cm,angle=270]{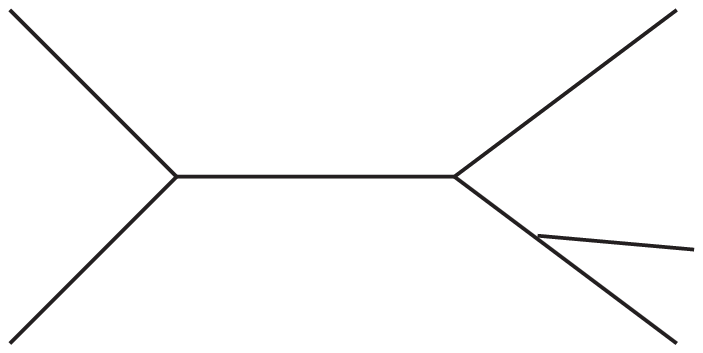}
 \end{center}
   \end{minipage} & \begin{minipage}{.5\textwidth}
\begin{center}
\includegraphics[width=2.5cm,angle=90]{emit.eps}
\end{center}
  \end{minipage}
\end{array}
\]
which, when they are squared up, generate collinear divergences in the cross-section. Adding the indistinguishable cross-sections of the above two sets of processes does not, however, lead to a cancellation of real and virtual collinear divergences.

A widespread response~\cite{sred:2007} to this is to say that an on-shell renormalisation scheme is inappropriate for a massless asymptotically free theory where the coupling is expected to be large for small momenta. Rather one should use an off-shell renormalisation scheme such as $\overline{\mathrm{MS}}$. In such a scheme the virtual loops are evaluated off-shell and thus do not contain collinear divergences.

Of course there are still collinear logarithms in the S-matrix. They now arise through the LSZ correction factor, which we denote by $\sqrt{\R}$ for each incoming and outgoing line, where from the residue of the propagator
\begin{equation}\label{phi3d60}
\R^{-1}=\left.\frac{d}{dp^2}S^{-1}\right\vert_{p^2=m^2}\,.
\end{equation}
This is also generated by multiplying the Green's functions by the ratio of renormalisation constants
\begin{equation}
\sqrt{\frac{Z^{\overline{\mathrm{MS}}}}{Z}}\,,
\end{equation}
for each external leg where $Z$ is the on-shell renormalisation constant and $Z^{\overline{\mathrm{MS}}}$ is in the $\overline{\mathrm{MS}}$-scheme. (Note that this ratio is UV finite.)

In our process this introduces four factors, one per external leg, which generate the following collinear logarithmic contribution to the cross-section
\begin{equation}\label{phi3d61}
\vert \mathcal{T_{\mathrm{virt}}}\vert^2= \vert \mathcal{T}_0\vert^2
\left[ \frac\alpha 3\ln(m^2)
\right]\,.
\end{equation}
Here $\vert \mathcal{T}_0\vert^2$ is the tree-level result, $\alpha=\lambda^2/(4\pi)^3$, and we are solely writing out collinear divergent terms.

In the spirit of the LN theorem, we now add the contribution of collinear emission from the two outgoing lines and collinear absorbtion on all incoming lines. These contributions are usually taken to be identical.\footnote{It is not immediately obvious that this should be the case and we feel that a satisfactory solution to the IR problem would not require such a strong requirement on the initial state. However, for the moment we will also make this assumption.} The four contributions, one per leg, of these real, collinear processes yield
\begin{equation}\label{phi3d62}
\vert \mathcal{T_{\mathrm{coll}}}\vert^2= \vert \mathcal{T}_0\vert^2
\left[ -\frac\alpha 3\ln(m^2)
\right]\,.
\end{equation}
It is apparent that the sum of the last two equations is collinear finite. We emphasise that this cancellation only occurs if one uses an off-shell scheme: in such schemes there is a one to one correspondence between the leg correction factors which underlies the above cancellation.  That this cancellation is scheme-dependent seems unsatisfactory as the LSZ correction factor is based upon mocking up the on-shell scheme, however, we will press on.

\medskip

Recall that the LN~theorem says that we should sum over \emph{all} possible degenerate processes and we  now argue that there are some further experimentally indistinguishable  processes. They are of two types. Firstly, as shown in the figure below:
\begin{center}
\includegraphics[width=2.5cm,angle=270]{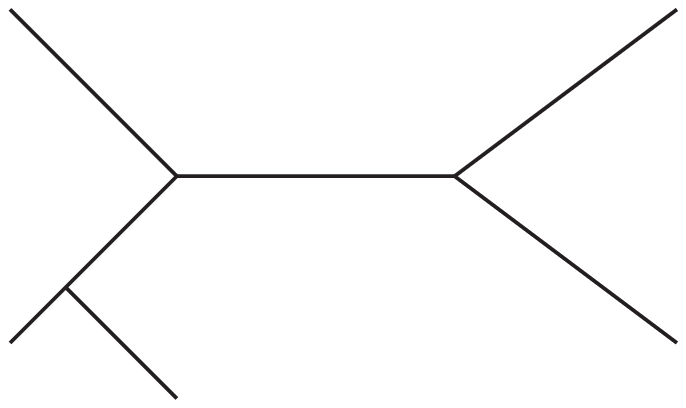}
\end{center}
if a real particle is emitted from an incoming line but collinear with one of the outgoing lines it would not be detected even if it was semi-hard. However, it will not generate a collinear divergence as the intermediate propagator in the diagram is not on-shell for these momenta. (There are similar harmless diagrams when a particle collinear to an incoming line is absorbed onto an outgoing line.)

However, the situation is different if a particle is emitted from an incoming line and is parallel to that same line or if an incoming particle parallel to an outgoing line is absorbed by that outgoing line.
\begin{center}
\includegraphics[width=2.5cm,angle=270]{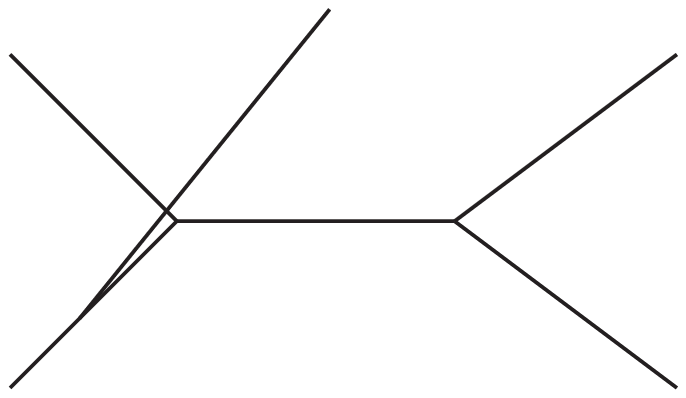}
\end{center}
This is because the intermediate propagator can now generate a collinear divergence. Such processes might be thought to be distinguishable, and indeed they are for semi-hard emission/absorbtion, but if such a particle is soft, i.e., has an energy below the experimental energy resolution $\Delta$, it will not be detected and so the cross-section for these processes should be added to those considered above. Due to the restriction on the energy, we will call this new class of processes \lq soft collinear\rq.

The result of calculating these four processes (soft collinear emission from incoming lines and absorbtion on outgoing lines) is another collinear divergence in the experimentally indistinguishable cross-section
\begin{equation}\label{phi3d63}
\vert \mathcal{T_{\mathrm{soft\ coll}}}\vert^2= \vert \mathcal{T}_0\vert^2
\left[ -\frac\alpha 2\left(\frac{\Delta^2}{\kb^2}\right)\ln(m^2)
\right]\,.
\end{equation}
The distinguishing feature of this contribution to the cross section obtained by summing over all degenerate states is the multiplicative power of
$\Delta$.\footnote{We recall from the introduction that setting the resolution to zero means that all cross sections will vanish.} The immediate question is what could cancel this singularity?

The above soft-collinear processes include either a (soft) final state particle emerging collinear with an incoming particle or a (soft) particle being absorbed collinear with an outgoing particle. Following the initial paper of Lee and Nauenberg~\cite{Lee:1964is} and see also~\cite{DeCalan:1972ya,Ito:1981nq,Muta:1981pe,Axelrod:1985yi,Akhoury:1997pb,%
Lavelle:2005bt,Lavelle:2006vv}, this suggests considering the interference of diagrams like
\begin{center}
\includegraphics[width=2.5cm,angle=270]{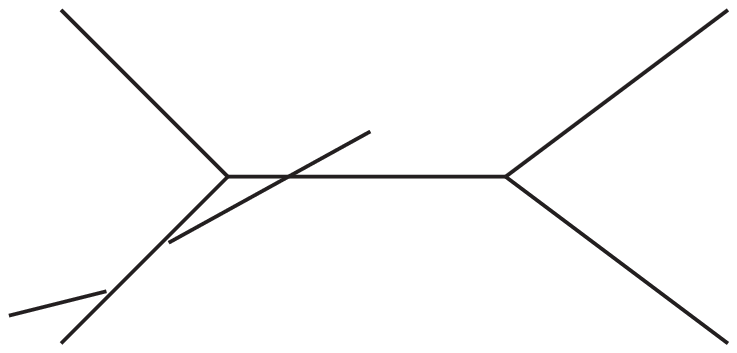}
\end{center}
i.e., emission and absorbtion on a line producing an interference contribution at the level of the cross-section with a process where a (soft) particle passes through the process collinear with that line.
\begin{center}
\includegraphics[width=2.5cm,angle=90]{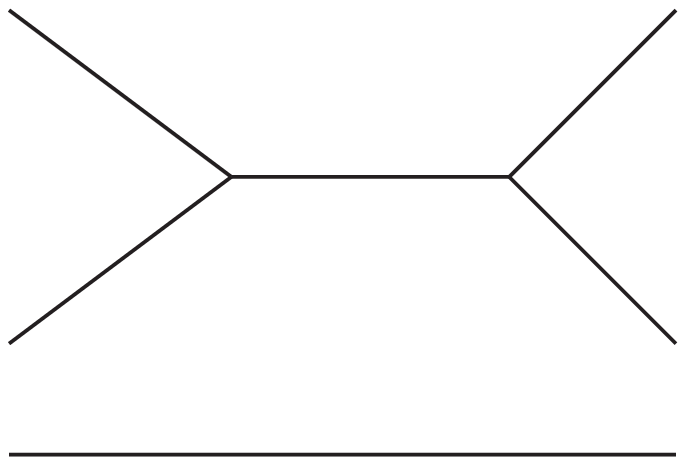}
\end{center}
It is known since the Lee-Nauenberg paper that such disconnected lines in S-matrix elements can produce a connected interference term in the cross-section (see Appendix~D of LN's paper and the discussion in~\cite{Lavelle:2005bt}). This enters the cross-section  at the same order in perturbation theory as the processes we considered above.

The result of calculating the connected interference of emission and absorbtion on each of the four lines with a disconnected line is an additional collinear logarithm
\begin{equation}\label{phi3d64}
\vert \mathcal{T_{\mathrm{disc\ soft\ coll}}}\vert^2= \vert \mathcal{T}_0\vert^2
\left[ 2\times \frac\alpha 2\left(\frac{\Delta^2}{\kb^2}\right)\ln(m^2)
\right]\,.
\end{equation}
The 2 inside the square brackets is a consequence of this being an interference contribution. It is evident that this does not cancel with (\ref{phi3d63}).

It should be noted that these divergences only arise in diagrams where the emission and absorbtion take place on the same line. Those diagrams with emission on one line and absorbtion on another such as
\begin{center}
\includegraphics[width=2.5cm,angle=90]{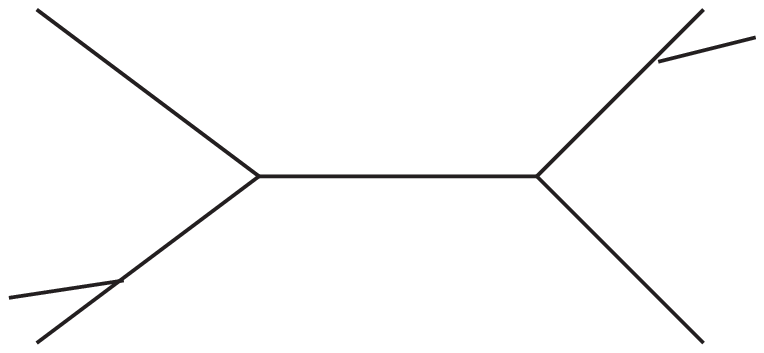}
\end{center}
do not generate collinear divergences in the cross-section when they interfere with the disconnected process.

The next set of diagrams we might consider are
\begin{center}
\includegraphics[width=2.5cm,angle=270]{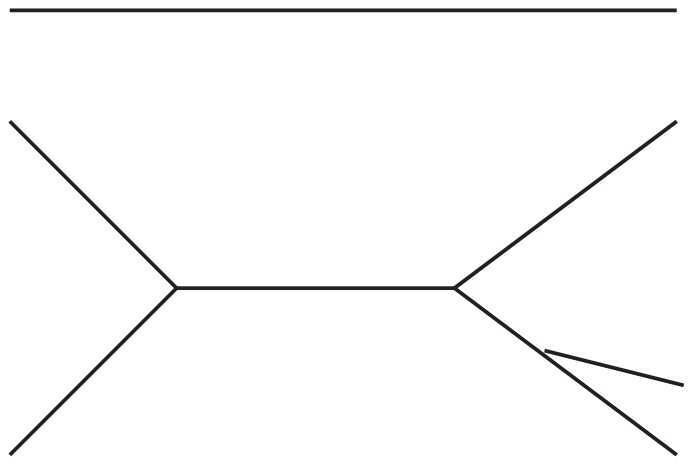}
\end{center}
i.e., the standard processes of emission on outgoing lines (and similarly for absorbtion on incoming lines) but now accompanied by a soft line going through. These diagrams can be squared and there is a connected contribution at the level of the cross-section.\footnote{We drop disconnected contributions to the cross-section as we expect they should be removed by normalisation.} This again contains collinear divergences
\begin{equation}\label{phi3d65}
\vert \mathcal{T_{\mathrm{coll\ with\ disc}}}\vert^2= \vert \mathcal{T}_0\vert^2
\left[ -\frac\alpha 2\left(\frac{\Delta^2}{\kb^2}\right)\ln(m^2)
\right]\,.
\end{equation}
The good news is that this appears to cancel our divergences, the bad news is that this last contribution is exactly equal to (\ref{phi3d63}). This implies that we could add processes like
\begin{center}
\includegraphics[width=2.5cm,angle=270]{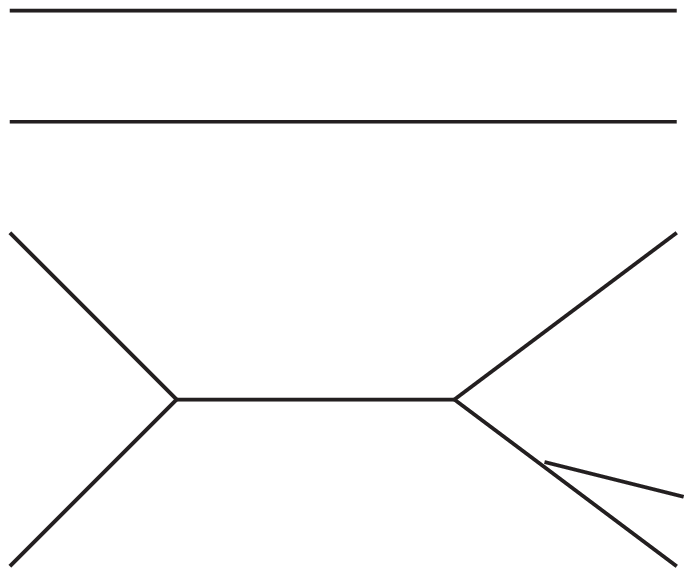}
\end{center}
i.e., with one extra disconnected line which, on squaring up, would generate the same divergent contribution to the experimentally distinguishable cross-section. There are in fact, at this fixed order of perturbation theory, infinitely many processes which each generate the same contribution to the cross-section. The series does not converge and hence we do not see a collinear finite cross section in this theory even in an off shell scheme.

\section*{The IR Structure of Massless QED}

The above issues are not exclusive to this simple model and neither are they restricted to off-shell schemes. In this section we will consider Coulomb scattering in on-shell massless QED (though retaining a small fermion mass as a collinear regulator) and investigate  infra-red divergent structures of the form considered above. One loop virtual diagrams yield soft divergences (apparent as $1/\epsilon$ factors in dimensional regularisation, $D=4+2\epsilon$), leading  ($\ln^2(m)$) and sub-leading collinear logarithms ($\ln(m)$) as well as mixed singularities ($\ln(m)\times1/\epsilon)$). The IR divergent terms in the  $F_1$ structure function (which is where the IR divergences reside) are
\begin{equation}\label{coul1}
2F_1=\frac{e^2}{4\pi^2}\left[-\frac1{\hat{\varepsilon}}\left(\!\ln\left(\!
\frac{\!Q^2}{m^2}\right)-1\right)
+\frac12\ln^2\!\left(\!\frac{\!Q^2}{m^2}\right)+\frac12\ln\!\left(\!\frac{\!Q^2}{m^2}\right)
\left(1-2\ln\!\left(\!\frac{\!Q^2}{\mu^2}\right)\right)\right]\,.
\end{equation}
Here $1/\hat{\varepsilon}=1/\varepsilon+\gamma-\ln4\pi$ and $\mu$ is a
dimensional regularisation mass scale, $Q^2=-q^2$ where $q$ is the energy-momentum transfer.

The BN trick says one should, at the level of probabilities, add the effects of emitting soft photons from the incoming and outgoing fermion lines. This cancels almost all the divergences leaving the tree-level cross-section multiplied by a sub-leading collinear logarithm
\begin{equation}\label{coul2}
\frac{e^2}{\pi^2}\ln\left(\frac Em\right)\left[\frac34-\ln\left(\frac E\Delta\right)
+\frac{\Delta}{2E}-\frac{\Delta^2}{8E^2}
\right]\,.
\end{equation}
Note that $E$ is the electron energy and $\Delta$ the energy resolution in the lab frame. (The virtual diagrams do not, of course, introduce any dependence on $\Delta$.)

Massless electrons are argued to be indistinguishable from such fermions accompanied by semi-hard photons inside a cone with angular resolution $\delta$. Hence, in the spirit of the BN trick, one considers the cross-section for the emission of semi-hard photons from the outgoing fermion line. This diagram generates
\begin{equation}\label{coul3}
-\frac12\frac{e^2}{\pi^2}\ln\left(\frac {E\delta}m\right)\left[\frac34-\ln\left(\frac E\Delta\right)
-\frac{\Delta}{E}+\frac{\Delta^2}{4E^2}
\right]\,.
\end{equation}
Adding the cross-section contributions in the last two equations cancels all of the terms proportional to powers of the energy resolution, $\Delta$, as we are now including emission with energies both below and above $\Delta$.  However, this diagram only generates  half of the remaining collinear divergences needed to cancel the singularities in (\ref{coul2}).

As we have repeatedly stressed, the LN argument says that one should add all possible degenerate processes. In particular, LN included the absorbtion of semi-hard photons on the incoming fermion line but not the absorbtion of soft photons on these lines. However, including only semi-hard absorbtion means imposing a non-zero lower limit on the emitted energy. Thus this reintroduces the singularities proportional to $\Delta$ and $\Delta^2$. One can try to remove these by including soft collinear absorbtion on to the initial line but this will also bring with it soft divergences. We conclude that the suggestion of treating soft divergences via the BN~trick and collinear ones via LN is untenable.

Following instead the fully inclusive LN approach for all IR~divergences, one should include all possible indistinguishable  processes: soft and semi-hard absorbtion and emission including interference effects from disconnected diagrams. It has been previously shown~\cite{Lavelle:2005bt,Lavelle:2006vv} that this yields a non-convergent series of soft divergences. However, the processes discussed earlier in this paper, viz.~soft emission from an incoming line and soft absorbtion onto an outgoing line, were not previously included. These yield additional collinear divergences multiplied by powers of the energy resolution.

\section*{Conclusions}

In this paper we have considered an asymptotically free theory, $\phi^3$ in six dimensions. This does not have soft divergences but does have the paradigm collinear divergences of QCD which are  important for jet formation. If one uses an off-shell renormalisation scheme, the virtual diagrams do not introduce collinear divergences but the leg correction factors make the S-matrix and cross-section collinearly divergent. Although these divergences from the leg corrections can be removed at the level of the cross-section by including emission from outgoing and absorbtion on incoming legs, these are not the only collinearly divergent degenerate processes.
We have here pointed out that there are other divergences which arise from soft collinear emission/absorbtion on incoming/outgoing lines. These processes yield collinear logarithms multiplied by powers of the energy resolution.

The origin of these divergent structures is that, for example, emission of collinear particles from an outgoing line should include both soft and semi-hard particles, i.e., one integrates over the energy resolution, $\Delta$, and there is no resolution dependence. However, as semi-hard collinear emission from an incoming line would be distinguishable, one should only include soft emission from these lines and this generates the $\Delta$ dependence.

We have pointed out that there are still more processes including disconnected lines which generate connected contributions at the level of the cross-section. However, due again to the masslessness of the fields, there are infinitely many diagrams at fixed order in perturbation theory and the series does not converge.

Such non-convergence has also been seen in QED and we have pointed out in this paper that soft emission/absorbtion also introduce collinear divergences multiplied by energy resolution in gauge theories.

This means that there are different sums of collinear divergences: those with and without a dependence on the energy resolution, $\Delta$.  They need to all cancel but it is not clear how they can in the standard frameworks.  Indeed, for simplicity we have kept to a single energy resolution above but a more realistic description of an experiment which may include different resolutions for incoming and outgoing particles makes such a cancellation much harder to see (some of the soft collinear divergent terms are multiplied by powers of a specific resolution, others by the lower resolution etc.).

If one recalls that the origin of infra-red divergences is that the asymptotic particles are not free~\cite{Dollard:1964,kulish:1970,Horan:1999ba}, the origin of the infra-red singularities may be recognised to be the use of free in and out states. Rather, we argue, one should use physical states~\cite{Bagan:1997su,Bagan:1999jk}. A minimal requirement in a gauge theory is local gauge invariance but this is not enough to specify the correct states~\cite{Dirac:1955ca,Bagan:1999jf}. We have seen in this paper that even in non-gauge theories other requirements are needed to determine the correct asymptotic states which can describe scattering in terms of an infra-red finite S-matrix and cross-section.

\bigskip\bigskip

\noindent \textbf{Acknowledgments:} TGS is grateful for the financial support of the Natural Sciences and Engineering Research Council of Canada (NSERC) and thanks the Plymouth high-energy physics group for their generous hospitality.

\end{document}